\documentclass[conference]{IEEEtran}

\AtBeginDocument{%
  \providecommand\BibTeX{{%
    \normalfont B\kern-0.5em{\scshape i\kern-0.25em b}\kern-0.8em\TeX}}}

\usepackage{amsmath,amssymb,amsfonts}
\usepackage{algorithmic}
\usepackage{graphicx}
\usepackage{textcomp}
\usepackage[hidelinks]{hyperref}
\usepackage{url}

\usepackage{xspace}
\usepackage{cleveref}
\usepackage{booktabs}
\usepackage{float}
\usepackage{circledsteps}
\usepackage{xcolor}
\usepackage{pdfpages}

\usepackage{tabularx}

\makeatletter
\let\MYcaption\@makecaption
\makeatother

\usepackage{subfig}

\makeatletter
\let\@makecaption\MYcaption
\makeatother

\usepackage{threeparttable}

\usepackage{graphicx}
\usepackage{etoolbox}

\definecolor{OCircle}{HTML}{AB73FF}
\definecolor{CCircle}{HTML}{3CC5F2}

\newcommand{\ovcircle}[1]{%
\begin{tikzpicture}[enum/.style={circle,draw=gray,very
    thin,fill=OCircle,text=white,inner sep=1pt},baseline=-3pt]
  \node (node 1) at (0,0) [enum] {\scriptsize #1};%
 \end{tikzpicture}}

\newcommand{\insertfig}{\includegraphics[width=\linewidth]{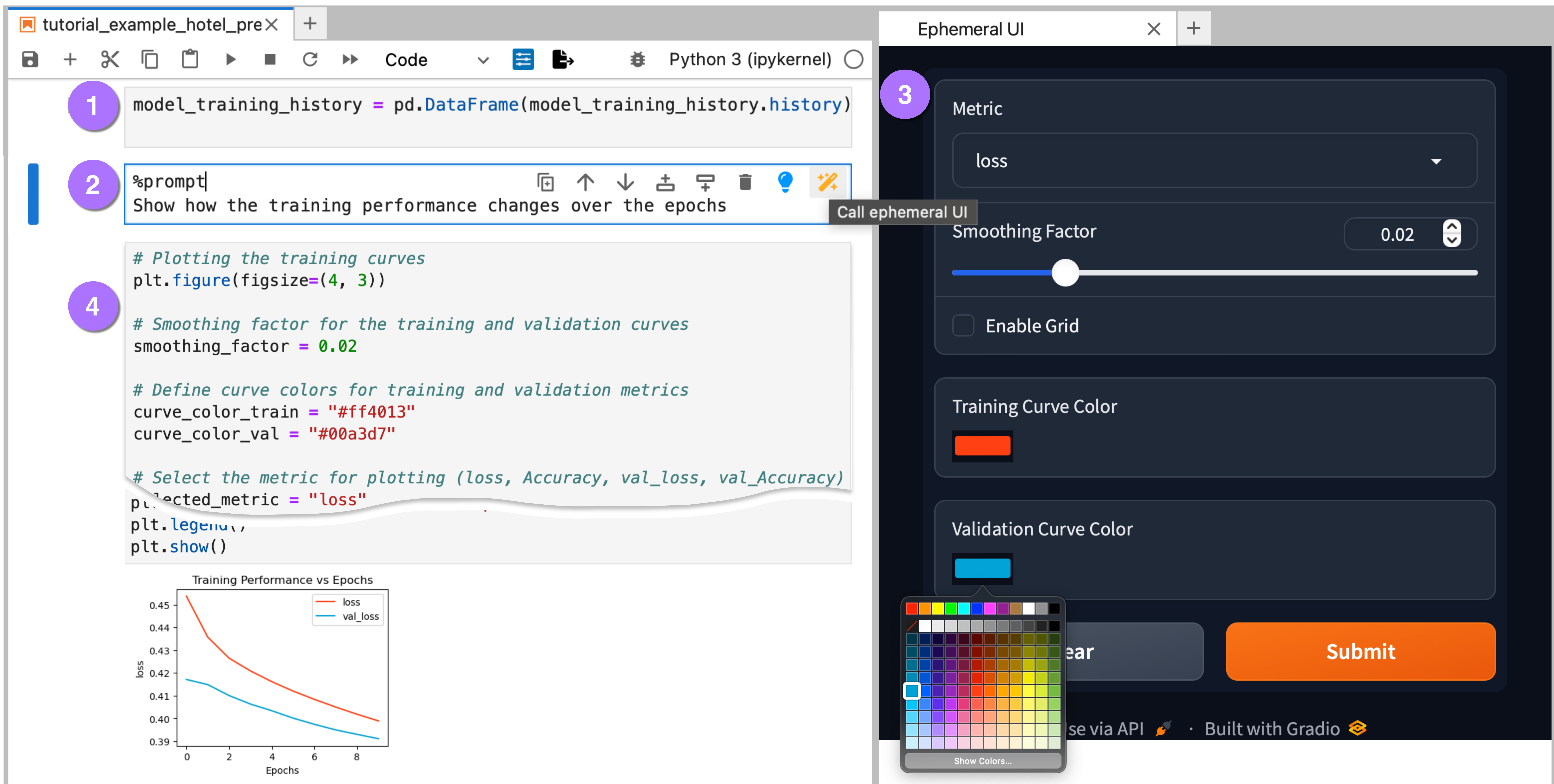}\captionof{figure}{\system is a JupyterLab extension prototype that offers LLM-generated UI elements to support programmers working with code in machine learning tutorials. With \system, a user working on an existing code cell in the notebook (\protect\ovcircle{1}) can enter a natural language request (\protect\ovcircle{2}), which triggers helpful ephemeral UI elements to be generated and displayed on a side panel (\protect\ovcircle{3}). The user can interact with the UI elements and get LLM-generated code based on their interaction with the UI (\protect\ovcircle{4}).}\addtocounter{figure}{-1}\label{fig:system-interface}}

\makeatletter
\apptocmd{\@maketitle}{\centering\insertfig}{}{}
\makeatother

\definecolor{mycolor}{HTML}{FFAC1C} 

\usepackage{tikz}

\usepackage{xcolor, listings, framed}
\definecolor{shadecolor}{gray}{0.95}
\newcommand{\system}{\textsc{Biscuit}\xspace}

\begin{document}

\title{\textsc{Biscuit}: Scaffolding LLM-Generated Code with Ephemeral UIs in Computational Notebooks}















\author{\IEEEauthorblockN{Ruijia Cheng}
\IEEEauthorblockA{Apple\\
rcheng23@apple.com}
\and
\IEEEauthorblockN{Titus Barik}
\IEEEauthorblockA{Apple\\
tbarik@apple.com}
\and
\IEEEauthorblockN{Alan Leung}
\IEEEauthorblockA{Apple\\
alleu@apple.com}
\and
\IEEEauthorblockN{Fred Hohman}
\IEEEauthorblockA{Apple\\
fredhohman@apple.com}
\and
\IEEEauthorblockN{Jeffrey Nichols}
\IEEEauthorblockA{Apple\\
jwnichols@apple.com}
}

\maketitle


\begin{abstract}
Programmers frequently engage with machine learning tutorials in computational notebooks and have been adopting code generation technologies based on large language models (LLMs). However, they encounter difficulties in understanding and working with code produced by LLMs. 
To mitigate these challenges, we introduce a novel workflow into computational notebooks that augments LLM-based code generation with an additional \textit{ephemeral UI} step, offering users UI scaffolds as an intermediate stage between user prompts and code generation. We present this workflow in \system, an extension for JupyterLab that provides users with ephemeral UIs generated by LLMs based on the context of their code and intentions, scaffolding users to understand, guide, and explore with LLM-generated code.
Through a user study where 10 novices used \system for machine learning tutorials, we found that \system offers users representations of code to aid their understanding, reduces the complexity of prompt engineering, and creates a playground for users to explore different variables and iterate on their ideas.

\end{abstract}

\section{Introduction}


Programmers delving into machine learning frequently work with interactive tutorials in computational notebooks, where they complete coding exercises with examples. The advent of large language models (LLMs) has introduced them to LLM-based code completion technologies such as GitHub Copilot,\footnote{\url{https://github.com/features/copilot}} OpenAI ChatGPT,\footnote{\url{https://chat.openai.com}} and Google Gemini.\footnote{\url{https://gemini.google.com}} Users interact with these tools by entering requests in natural language or directly in code, and the LLMs respond with high-quality code snippets accompanied by comments and explanations catered to their specific context and intentions~\cite{madi2023how, kazemitabaar_studying_2023}. However, 
this mode of interaction can present challenges for users in understanding and exploring beyond the code produced by LLMs---for example, in comprehending the rationale behind certain syntax and logic in code generated by LLMs~\cite{vaithilingam2022expectation, bird_taking_2023}.
Users may also overrely and overtrust code generated by LLMs, overlooking potential issues and alternative solutions~\cite{wang2023investigating, johnson2023make}.




We introduce a \textit{UI-centric} approach for users to interact with LLMs in code generation. Instead of directly generating code based on user requests, our approach offers dynamically-generated UIs as an additional layer of scaffolds between users' natural language requests and code generation. Inspired by prior work on UI-based scaffolds in programming and the emerging body of literature on LLM-generated UIs~\cite{vaithilingam2024dynavis, dibia2023lida}, we devise a workflow of \textit{ephemeral UIs}---UI elements that are dynamically generated by LLMs and contextually integrated with the code context and user requests. This workflow allows users to interact with UI-based scaffolds in code generation, facilitating code comprehension and exploration. 

Specifically, we present this workflow in \system (Building Interactive Scaffolding for Code Understanding In Tutorials). As interactive tutorials for machine learning are commonly hosted in computational notebooks, we implemented \system as a JupyterLab extension to offer in-context support.
As shown in \Cref{fig:system-interface}, with \system, users can trigger ephemeral UIs through natural language requests. Useful UI elements are generated by the underlying LLMs based on the code context and user request. Users can interact with these UI elements, leading an LLM to generate and inject code into the notebook. 

We conducted 10 user study sessions where programmers who are novices in machine learning engaged with \system to work with an interactive tutorial in JupyterLab. We found that \system supported users for understanding LLM-generated code by surfacing key variables that users can experiment with. \system allowed users to guide code generation, as it reduces the effort of detailed prompting and offers users an intuitive interface to customize the generated code. \system also creates a playground for iteration and offered users inspirations for alternative implementations. Despite \system introducing an additional step where users interact with UIs in the code generation process, users found that \system in general enhances the efficiency of their work with tutorials. 


We make the following contributions: 1) A novel workflow of ephemeral UIs that offers dynamically generated scaffolds for code comprehension and exploration catered to users' intention and code context. 2) An LLM-based implementation of the workflow as a JupyterLab extension prototype, \system. 3) Empirical findings that demonstrate how \system supports programmers in machine learning to understand, guide, explore, and efficiently work with code in machine learning tutorials. 

%

\section{Example Usage Scenario for \system}
\label{sec:usage_scenario}

\begin{figure*}[ht]
\centering
\addtocounter{figure}{1}
\includegraphics[width=0.85\textwidth]{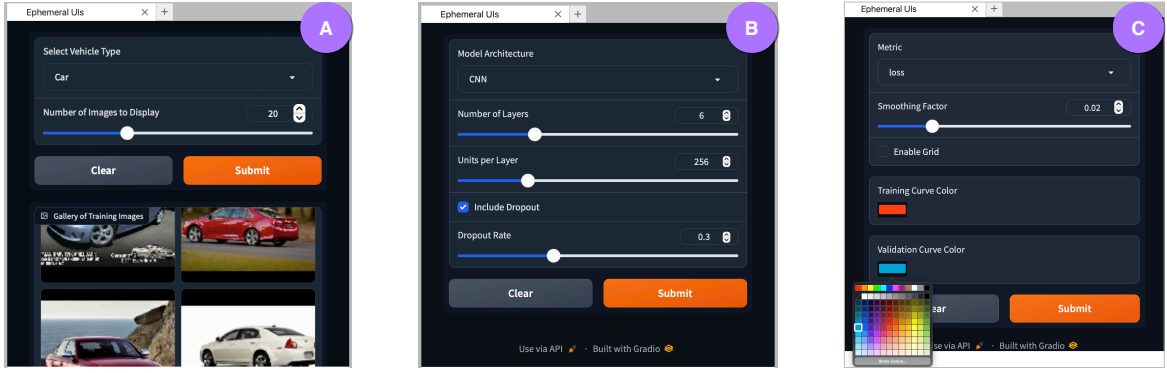}
\caption{Ephemeral UIs generated by \system in the example usage scenario described in \Cref{sec:usage_scenario}. \protect\ovcircle{A}: generated UI for sampling data in an image dataset; \protect\ovcircle{B}: generated UI to support model customization; \protect\ovcircle{C}: generated UI to support visualization.}
\label{fig:use_scenario}
\end{figure*}


Amy, a software developer new to Python and machine learning, is working with a machine learning tutorial on image classification in JupyterLab. She installs \system as an extension to the JupyterLab environment.

Amy first invokes \system by entering a request in natural language: ``Show me a sample of the dataset images.'' \system generates and displays an ephemeral UI, including a dropdown menu of labels in the dataset and a slider, to determine the size of the sample (\Cref{fig:use_scenario} \ovcircle{A}). Amy selects a label from the dropdown menu and adjusts the slider to display 20 images in the image gallery area in the UI. Amy views the sampled images and selects another label for a new sample. Getting a basic sense of the data, Amy progresses with the tutorial. 

Amy now wants to customize a model defined in the tutorial but she is unsure how to modify the code. She enters ``What are some other ways to construct the model'' and gets a new ephemeral UI with a dropdown menu of model architecture, sliders for number of layers and units per layer, and UI elements for dropout layers (\Cref{fig:use_scenario} \ovcircle{B}). She selects values from the UI and clicks the submit button. A new block of code that defines a model layer structure based on Amy's selections in the UI is generated and added to the notebook. 

Going forward, Amy wants to visualize the training performance and asks: ``Visualize the training performance.'' She gets a UI for constructing visualizations, including a dropdown menu for training metrics, color pickers for the curves, and elements to specify other features in the plot (\Cref{fig:use_scenario} \ovcircle{C}). Amy makes a selection, uses it to generate code, and runs the code to produce a line chart with the selected colors. Amy then changes her selections in the UI and generates another visualization on a different training metric to understand how different metrics change during the training process.

\section{Related Work}

\subsection{Dynamically generated intelligent UIs}
The HCI community has been actively researching dynamic user interfaces that are tailored to usage context in various domains. Early systems in accessibility and ubiquitous computing leverage user traces and inputs to recommend user interfaces. For example, SUPPLE~\cite{gajos2004supple} automatically adjusts user interfaces to fit the constraints of computational devices and customizes interface renditions based on individual user's usage patterns; SUPPLE++~\cite{gajos2007automatically} leverages assessments of user's motor capabilities and uses optimization algorithms to recommend personalized interfaces. An important rationale of these dynamic UIs is to mitigate the challenges users face when navigating complex interfaces, especially in systems that involve multiple connected appliances. Dynamically generated interfaces are thus designed to align with user's habits, preferences, and context-specific needs. For example, PUC, a personal universal controller~\cite{nichols2003personal} automatically generates graphical and speech interfaces and can reduce task time and errors compared to the static interfaces provided by manufacturers. This approach has been further advanced in systems like UNIFORM~\cite{nichols2006uniform} and HUDDLE~\cite{nichols2006huddle}, which continue to build on automatically generated and context-aware interface design in remote control.

In programming, tools have been developed to dynamically generate user interfaces tailored to user needs. For instance, Heer et al.~\cite{heer2008generalized} introduce a method for generating dynamic UIs through query relaxation, enabling users to generalize their selections. Mavo~\cite{verou2016mavo} empowers users to create interactive HTML pages by adding special attributes and allowing them to interact with editing widgets that are recommended based on these attributes. Bespoke~\cite{vaithilingam2019bespoke} offers a way for users to create custom graphical user interfaces for command-line applications, enhancing accessibility and usability. NL2INTERFACE~\cite{chen2022nl2interface} synthesizes SQL queries from natural language commands and generates a UI for users to edit the parameters or variables in the SQL query. Inspired by these advancements, we integrate user interfaces that are specifically tailored to the programming context within a user's programming workflow.

The recent advancement of LLMs has fueled interest in using these models to generate UI elements. These tools allow users to articulate their intentions in natural language to create UI components. For instance, LIDA~\cite{dibia2023lida} offers an interface with which users can create data visualizations directly from natural language. DynaVis~\cite{vaithilingam2024dynavis} uses an LLM to synthesize interactive visualizations with dynamically generated UI widgets that are customized to the visualizations and enable direct manipulation of visualization properties. In alignment with these new developments, our system leverages LLMs to interpret user intentions and accordingly generate UI components. Our approach extends beyond the domain of visualizations with the goal of making the interaction with machine learning tutorials more intuitive and effective.

\subsection{Tutoring scaffolds in machine learning}

Prior research has explored tutoring systems for machine learning and programming more broadly. To scaffold novices, graphical UIs have been developed to allow for direct manipulation in programming systems. For example, Online Python Tutor offers an interface for displaying the runtime state of data structures, enhancing learners' understanding of program execution~\cite{guo2013online}. Tools such as Wrangler~\cite{kandel2011wrangler}, Wrex~\cite{drosos2020wrex}, and Unravel~\cite{shrestha2021unravel} offer interactive components for users to directly engage with data and see the immediate impact of their actions in data tables. In the context of computational notebooks, Mage offers an API that allows users to construct customized UIs themselves~\cite{kery2020mage}, facilitating bidirectional editing of data.

Beyond UI scaffolds, intelligent tutoring systems offer scaffolding through additional documentation and resources. 
An emphasis of these tools is to provide explanations or comprehension support to aid users in their programming workflows. For instance, Tutorons~\cite{head2015tutorons} offers context-relevant, on-demand explanations and demonstrations of online code snippets, reducing the need for learners to consult external documentation. Recent tools have introduced LLM-powered explanations adapted to the user's specific programming contexts. Ivie~\cite{yan2024ivie}, for example, provides in-situ and inline explanations for code generated by LLMs, enhancing users' understanding of the generated code. Furthermore, Nam et al.~\cite{nam2024using} introduce an IDE plugin that utilizes LLMs to elucidate code sections through highlighting, key domain-specific terms, and usage examples for APIs.

Inspired by these LLM-powered tutoring systems, our system utilizes LLMs to enhance users' understanding of code. Distinct from existing systems that directly provide code explanations, our system offers an interactive experience involving user prompts and dynamically generated UIs that scaffold reflective practices in the comprehension process~\cite{schon2017reflective}. 
Moreover, our approach extends beyond mere code comprehension. Drawing inspiration from the design space of LLM-based features in computational notebooks~\cite{mcnutt2023ai}, our system aims to support the user's programming workflow in notebooks by offering LLM-generated UI support.

\section{System}
\label{sec:system}


The design goals of \system are informed by existing research and formative discussions. 
\system is a prototype implemented as an extension for JupyterLab. 
The backend of the extension uses OpenAI's Chat Completion API\footnote{\url{https://platform.openai.com/docs/guides/text-generation/chat-completions-api}} to interpret user intent and to generate code. 
Following the approaches in Ferreira et al.~\cite{ferreiraexamples}, the LLMs in our system leverage the Gradio API\footnote{\url{https://www.gradio.app}}---a Python library that offers a catalog of common and useful UI components for machine learning---to implement interactive elements as part of machine learning tutorials.


\subsection{Design goals}
\label{sec:system_design_motivations}
The design of our system is informed by existing research on intelligent interfaces and tutoring systems, as well as formative discussions with five experts from our organization, including two machine learning researchers, one software engineer, and two hardware engineers who have experience with LLM-based code generation tools and machine learning tutorials in computational notebooks. We discussed their usage of code generation tools as well as any suggestions for improving the code generation experience in machine learning workflows.
We describe our design goals (DGs) as follows:







\textbf{DG1: Scaffolding understanding with UI-centric support.} 
Our formative discussions surface the potential for UI scaffolding in code generation. For instance, one expert shared that engineers in their team 
wished for UIs with attributes of a given machine learning method to allow them to quickly grasp and understand the various properties involved. This feedback echos decades of research supporting the use of UIs in programming, which suggests that UIs can reduce the entry barrier and provide means for direct manipulation~\cite{kandel2011wrangler, drosos2020wrex}. In our design, we introduce dynamically generated UI elements that correspond to code segments that users can manipulate. 

\textbf{DG2: Facilitating user guidance in code generation.} 
While users value the flexibility of LLMs in code generation, they also desire control over the process. We learned that although users could have trouble understanding and working with generated code, they nevertheless wanted to maintain the generative capabilities of LLMs to enable serendipitous discoveries of new methods and techniques. This reflects ongoing discussions about scaffolding users to guide code generation~\cite{barkeGroundedCopilotHow2022, jiang_discovering_2022}. Our design introduces ephemeral UIs that allow users to specify details in the code to be generated and enables users to view and fully edit the code produced by LLMs.

\textbf{DG3: Empowering users to explore and expand code examples.} 
Users consider it beneficial to investigate alternative approaches to the provided examples in machine learning tutorials, echoing existing research that demonstrates the importance of offering a wide array of examples for programming concepts~\cite{cheng2022how,ni2021recode, drosos2020wrex}. Our system is designed to aid users in exploration with code, where the UI elements correspond to parts in the code examples that users can change. Our system also allows users to generate different versions of code by making selections in the UIs. 

\textbf{DG4: Offering in-context scaffolds.} 
It can be disruptive for users to switch to a different application outside of the notebook to use LLMs to generate code or explanations. Therefore, systems such as Tutorons~\cite{head2015tutorons} and Mallard~\cite{xiong2019mallard} provide in-situ explanations and demonstrations within the webpage a user is working on, and recent advancements in programming support tools often feature inline assistance within the IDEs~\cite{nam2024using, yan2024ivie}. In light of this, our design integrates UI scaffolds into the JupyterLab interface from the user's code context.

\subsection{Functionalities}
\begin{figure}[b]
  \centering
  \includegraphics[width=0.8\linewidth]{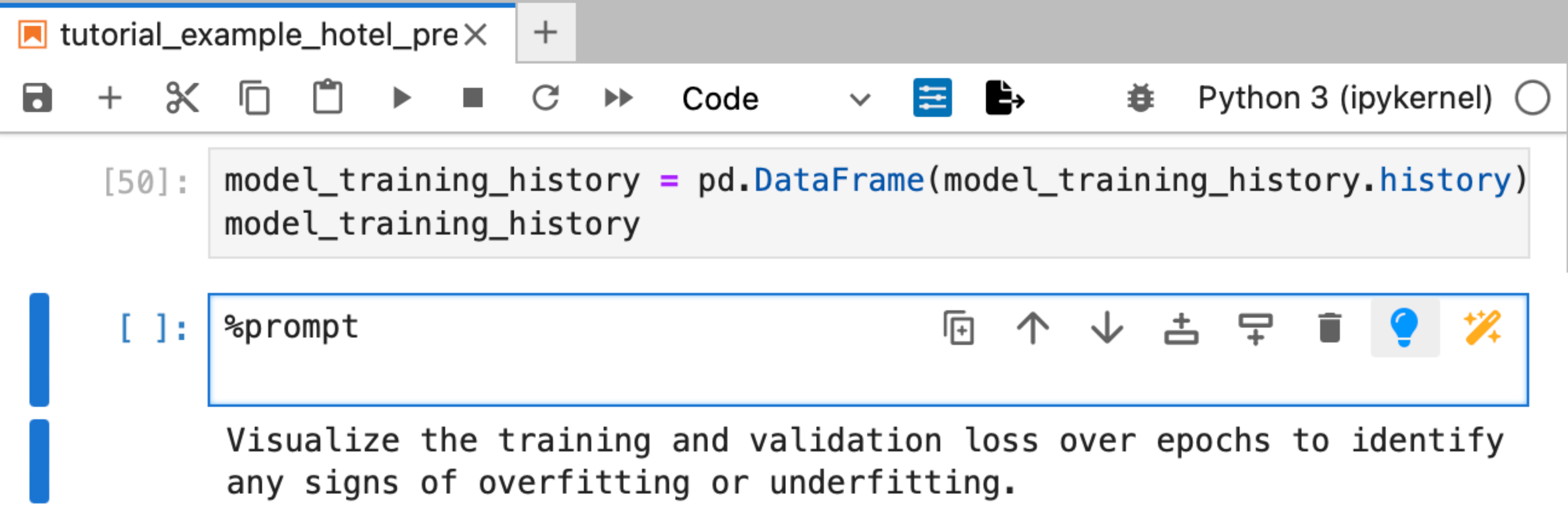}
  \caption{Helper feature of Prompt Suggestion.}
  \label{fig:system-helpers}
\end{figure}

\begin{figure*}[ht]
  \centering
  \includegraphics[width=0.8\linewidth,alt={}]{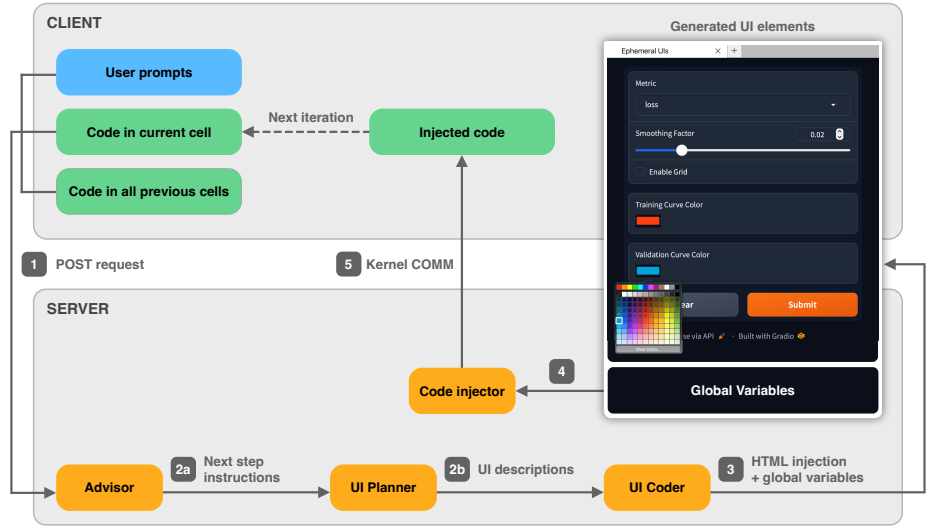}
  \caption{System implementation of Ephemeral UI.}
  \label{fig:system-flow}
\end{figure*}
The interface of \system consists of Code Context, User Request, Ephemeral UI, and Code Injection (\Cref{fig:system-interface}). 

\paragraph{Code Context}
Code context refers to the existing code in a cell based on which the user wishes to request an ephemeral UI. In line with \textbf{DG3}, users can extend and build on any code examples in tutorials. 
For instance, as depicted in \Cref{fig:system-interface}, the code context is in the cell highlighted in \ovcircle{1}, which contains code for loading a dataframe of training history. 

\paragraph{User Request}
Following \textbf{DG2}, users can guide code generation through ephemeral UIs via natural language requests that articulate their intentions in relation to the code context. Users can insert an empty cell beneath any code cell in the notebook, initiate it with the string ``\texttt{\%prompt}'', and then type in their natural language request. For instance, as illustrated in \Cref{fig:system-interface}, the user's request on the code context in \ovcircle{1} is showcased in \ovcircle{2}, with the instruction: ``Show how the training performance changes over the epochs.''

When needing help devising a request, the user can activate the Prompt Suggestion feature as shown in \Cref{fig:system-helpers}. In any of the prompt cells (code cells that start with the string ``\texttt{\%prompt}''), the user can click the blue ``light bulb'' button on the cell tool bar. A suggested natural language request will be generated by an LLM agent (independent to the ones in the Ephemeral UI) based on the user's original prompt (if any) and the code context. The suggested prompt will be printed in the output area of the prompt cell and guides the user to explore and expand their thinking around the current code. 

\paragraph{Ephemeral UI}
The user request prompts the system to provide relevant UIs (\ovcircle{3}) that assist code generation, reflecting principles outlined in \textbf{DG1} and \textbf{DG2}. 
After providing the natural language request, the user can generate an ephemeral UI by clicking the orange ``magic wand'' button in the cell. Following \textbf{DG4}, UI elements are presented in-context with the tutorial the user is engaged with, appearing in a panel on the right side of the JupyterLab interface. The generation process takes into account the prompt and the code present in the code cell immediately preceding the prompt cell, while using all preceding code within the notebook as supplementary context. \Cref{fig:system-interface} shows an example where the user interacts with a series of UI elements generated by the LLMs to help them visualize the training performance.

\paragraph{Code Injection}
Following \textbf{DG2}, our system allows users to guide the code generation process using UI elements. Upon submission, an LLM agent creates code based on the user's selections within the UI elements, the code context, and the user's prompt. This newly generated code is then inserted into a new code cell positioned directly beneath the user request. Users can edit and execute the injected code just as they would with any other code cell within their environment. For instance, as illustrated in \Cref{fig:system-interface} \ovcircle{4}, the underlying LLM produces code that visualizes both training and validation losses, reflecting the choices made by the user through the ephemeral UI.
The user can make new selections within the UI elements to generate additional variations of code. Furthermore, the user can re-click the magic wand button or enter a new prompt to generate a fresh set of UI elements. These new elements will replace the existing ones in the panel.



\subsection{Implementation of \system}

\Cref{fig:system-flow} illustrates the implementation details of \system. When the user activates the ephemeral UI by clicking the magic wand button in a prompt cell, the user request, code in the current cell, and code in all previous cells above the current cell will be sent from the client to the server through an HTTP POST request, as in \ovcircle{1}. This information will be used by an underlying LLM agent ``Advisor,'' which makes instructions on a concrete next step based on a user request.

The instruction will be passed to a second LLM agent, the ``UI Planner,'' as illustrated by \ovcircle{2a}. The UI Planner will describe a list of UI elements helpful for the suggested next step. The output of the UI Planner is a JSON data object with a name label and a description of the UI element, and a unique identification number.

The output of the UI Planner will be passed to a third LLM agent, the ``UI Coder,'' as shown in \ovcircle{2b}. The UI Coder will generate Python code to implement the UI elements. Specifically, it generates two snippets of Python code. It will generate a snippet of code to declare a series of global variables corresponding to each of the UI elements described in \ovcircle{2b} to store and update values based on user interaction. It will also generate a series of methods that implement the UI elements using the Gradio API. These code snippets will then be sent to the Kernel to be executed, shown by \ovcircle{3}.

After execution in the Kernel, a generated HTML string is injected into an empty panel created on the client JupyterLab interface and rendered as UI elements. Users can interact with the UI elements in the client, and in this process, the global variables corresponding to the UI elements will be updated in real-time to the user selected values.

Once the user clicks the ``Submit'' button, a fourth LLM agent, the ``Code Injector,'' will generate code based on the values of the global variables (\ovcircle{4}) as well as the earlier user request and code context given as input. The generated code will be injected into the cell content in a new code cell in the client through Jupyter's Kernel COMM protocol, as depicted by \ovcircle{5}. This process can repeat as the user enters new prompts and requests another ephemeral UI on the newly added code cell.

All four LLM agents are based on OpenAI’s Chat Completion API: the ``Advisor,'' ``UI Planner,'' and ``UI Coder'' use the \texttt{GPT-4-turbo-0125} model, and the ``Code Injector'' uses the \texttt{GPT-3.5-turbo} model. 

\subsection{Limitations of the prototype}
\system is a prototype that demonstrates the workflow of UI-centric code generation in computational notebooks. Designed to illustrate the concept of ephemeral UIs, it has inherent limitations. The generation of ephemeral UIs relies on the quality of the underlying GPT models from OpenAI. Occasionally, the ``UI Planner'' may produce an empty string, leading the client interface to show an empty UI. Model hallucination sometimes results in code that fails to compile. We did not invest in implementing any mechanism for automatic error correction.
Despite these challenges, the prototype acts as an useful instrument in our user study.
In scenarios where issues arise, users can simply reclick the magic wand button to make a new request and regenerate the UI. 

\section{User Study}


To understand how users' experiences with our system align with the four Design Goals outlined in \Cref{sec:system_design_motivations}, we conducted a user study with 10 participants in 1-hour, 1-to-1 interview sessions over video conference.



\subsection{Study procedures} 
\label{sec:study_procedures}

\subsubsection{Onboarding (5 minutes)}
The study begins with a 5-minute onboarding session where a researcher presents a slide deck to familiarize the participant with the system's features.

\subsubsection{Working with machine learning tutorials using \system (45 minutes)}
The participant is then asked to work with one of two randomly assigned interactive machine learning tutorials in JupyterLab using \system. In the tutorials, participants were instructed to opt in to the ephemeral UIs by entering a natural language request whenever they felt they would need the assistance.

The tutorial acts as a design probe in the study to bring ephemeral UIs to the forefront. Participants are instructed to engage actively with the tutorials by running and editing the provided code to answer questions and complete tasks. The specific two tutorials that we used in the study are adapted from public machine learning tutorials available on Kaggle's Learning module\footnote{\url{https://www.kaggle.com}}---including a tutorial about building a binary classification model on hotel cancellation using a tabular dataset of hotel reservation features (Binary Classification),\footnote{Adapted from the \textit{Exercise: Binary Classification} notebook~\cite{binary_classifier_tutorial}.} and a tutorial about building an image classifier on pictures of cars and trucks (Image Classification).\footnote{Adapted from the \textit{Exercise: The Convolutional Classifier} notebook~\cite{image_classifier_tutorial}.}  
Both tutorials are written in Python, contain common data types and libraries in machine learning, and take around 45 minutes to complete based on pilot studies in our team.
In preparing the tutorials, we kept the unmodified datasets and most of the content in both tutorials, while removing any code snippets that are specific to the Kaggle environment (for example, code for checking answers). We also added additional text instructions to prompt users to customize functions and models beyond the code examples. 

Each participant was randomly assigned to one of the two tutorials to avoid fatigue in the sessions (5 participants working with the Binary Classification tutorial and 5 with the Image Classification tutorial). Participants interacted with \system---installed on the researcher's device through the remote device control feature of the video conference software---to reduce the burden of installing JupyterLab and the extension.  
During the sessions, participants were encouraged to think-aloud and verbally describe what they were doing and their reactions to different ephemeral UIs and code generated by the LLM.

\subsubsection{End of session interview (10 minutes)}
In the end, participants were interviewed on their experience using the prototype and opportunities for future improvement. 
To investigate users' experiences and their perceptions of how well our system achieves each of the Design Goals, we asked participants a series of Likert scale questions based on our Design Goals (\Cref{sec:system_design_motivations}) and probed for the rationales for their ratings. Specifically, we asked participants to rate their agreement on the following statements about \system. 
\emph{Understanding code (\textbf{DG1}):} The system is useful in helping me understand the code examples in the tutorial and the code generated by the LLM.
\emph{Guiding code generation (\textbf{DG2}):} The system allows me to guide the code generation process.
\emph{Exploring code (\textbf{DG3}):} The system helps me explore beyond the code examples in the tutorial.
\emph{Efficiency with tutorials (\textbf{DG4}):} The system helps me work efficiently through the tutorial.
\emph{Overall usefulness:} Overall, the system is useful as a tool for me to work with machine learning tutorials.



\subsection{Participant recruitment}
\label{sec:method_participants}
\begin{table}[]
\centering
\begin{threeparttable}
\caption{User Study Participants\label{tab:participants}}
\begin{tabularx}{\columnwidth}{lXXl}
\toprule
\textbf{ID}  & \textbf{Job Title}           & \textbf{Python Level} & \textbf{Tutorial}\tnote{1}\\
\midrule
P1  & Research Scientist  & Experienced       & BC \\
P2  & Software Engineer   & Experienced       & BC \\
P3  & Software Engineer   & Intermediate      & IC  \\
P4  & Software Engineer   & Intermediate      & IC  \\
P5  & Software Engineer   & Novice            & BC \\
P6  & Software Engineer   & Intermediate      & IC  \\
P7  & Hardware Engineer   & Novice            & IC  \\
P8  & Research Engineer   & Experienced       & BC \\
P9  & Software Engineer   & Experienced       & IC  \\
P10 & Hardware Engineer   & Intermediate      & BC \\ \bottomrule
\end{tabularx}
\begin{tablenotes}
\item[1] BC: Binary Classification tutorial, IC: Image Classification tutorial
\end{tablenotes}
\end{threeparttable}
\end{table}

We recruited 10 participants (4 female, 6 male) from our organization. To recruit participants, we advertised our user study on our internal channels.
We selected a subset of 10 participants who had a range of existing experience levels with Python and had used JupyterLab or Jupyter Notebook before,
with a balance of gender identities and job titles in the organization.  
We selected participants who reported having no experience or were novices in all the machine learning libraries used in the tutorials, and having previous experience using LLMs in programming. Participants were compensated with a \$12 meal voucher on completion of the study. The profile of our participants can be found in \Cref{tab:participants}.

\subsection{Data collection and analysis}
We collected the audio and video recordings of the sessions and 
conducted qualitative analysis on the recorded transcripts, applying inductive coding to categorize participants' experiences with \system with respect to the four Design Goals.

\section{Results}
\begin{figure}[h]
  \centering
  \includegraphics[width=0.8\linewidth]{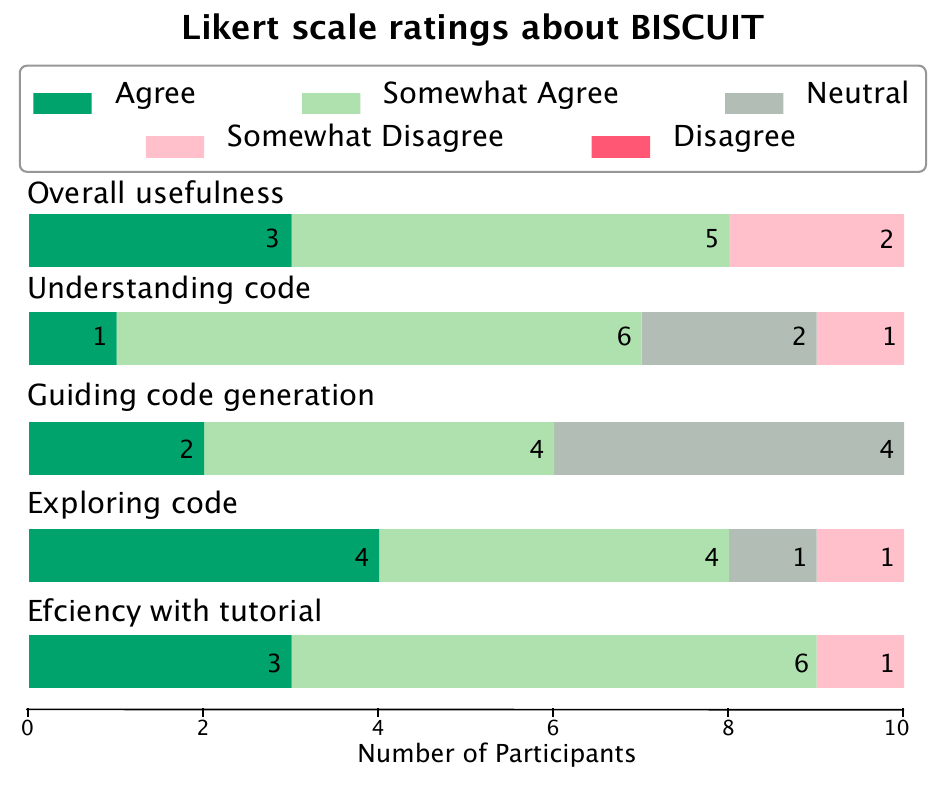}
  \caption{Results of the Likert scale questions from the user study.}
  \label{fig:likert_scale}
\end{figure}

Participants found \system helpful in augmenting LLM-based code generation. Eight out of 10 participants in our user study found \system to be a useful tool for them to work with machine learning tutorials (\Cref{fig:likert_scale}). 
The subsequent sections are organized according to users' perceptions of how our system achieves each of the Design Goals in helping them understand, explore, guide, and enhance their efficiency working with LLM-generated code. In each section, we first report the Likert scale results then describe the qualitative findings.

\subsection{Understanding code}
Most participants found \system helpful for aiding them to understand code that they were working on in the tutorial. \Cref{fig:likert_scale} shows 7 out of 10 participants either agree or somewhat agree that \system supports them in understanding code generated by the LLM. Participants highlighted that the ephemeral UIs supported their understanding of both LLM-generated code and existing code examples in tutorials by providing representations of code coupled with support for empirical experimentation, which they referred to as ``learning by doing.'' At the same time, participants found a lack of explanations in some of the ephemeral UIs and offered suggestions for opportunities to enhance these UIs with additional explanations to further augment their comprehension.


\subsubsection{Offering visual representation of code}
Participants found \system helpful with code comprehension by providing novices with a ``visual representation and visual playground'' (P5) of the code.
For example, in the part of the Binary Classifier tutorial where users were asked to complete the \texttt{model.compile()} function with an optimizer and performance metrics, P1, who had never programmed model training before, was initially not sure how to compile the model.
In the ephemeral UI generated based on her request to ``compile the model with optimizer and metrics,'' P1 was able to select ``Adam'' as the optimizer, loss as the metric, and learning rate of 0.05. P1 was then able to understand the function and the different properties.

P5 described a similar experience where an ephemeral UI (\Cref{fig:user_study_examples} \ovcircle{C}) helped him break down the tutorial code to process missing values in the data: ``I like the split screen view where I have the code on the left and the UI on the right. I can see the connection that I made here'' (P5).
The labels generated as part of the UIs helped users ``connect the dots'' (P5) between machine learning concepts and the implementation in code. For example, P3 interacted with a slider labeled ``First Layer Neurons'' (\Cref{fig:user_study_examples} \ovcircle{D}) and realized that it corresponded to the numbers defined in the example code: ``in this code I had no idea these numbers are the number of neutrons---and in the UI I am getting a visual hint.'' In summary, ephemeral UIs functioned as a visual representation of code, transforming code into accessible and comprehensible segments.

\subsubsection{Enabling ``learning by doing''}
Participants reported that they were able to achieve a deeper understanding of machine learning methods from experimenting in the UI and reading the corresponding code---what they referred as ``learning by doing'':
``You can learn more about the code if you're changing the elements and then you see some parts of the code change. So the learning comes from doing it, by messing with the UI'' (P9).
For example, P4 was trying to change a model structure and found the UIs helped him make meaningful changes to code: ``
Especially when there's existing code, sometimes you don't know what you can change without breaking it. You get a slider and you can just move it.''
According to him, the UI felt ``less scary'' compared to direct code manipulation. While he needed to understand the details of code in order to edit it, the UI provided a supportive environment that encouraged experimentation: ``something about code is definitive and fragile. The UI is easily undone and wants to be interacted with and is also easy to understand'' (P4).
As learning to program involves experimentation and tinkering, ephemeral UIs facilitate this process by offering easy and interactive options.



\subsubsection{Opportunity---supplementing ephemeral UIs with additional explanation}
Participants wanted explanations integrated with some of the UI elements in the ephemeral UIs to assist their learning and selection in context.
For example, P1, who was unfamiliar with the mechanisms of the different options of activation functions in a dropdown menu, desired more details: ``I was expecting a little bit of explanation for the options, kind of integrated with the explanations into the UIs. There were a lot of options. I feel like if there was a brief explanation, it would have been pretty helpful.''
Around more comprehensive explanations within ephemeral UIs, participants brought up ideas for future enhancement---such as integrating API documentation directly into the UI explanations and providing users with access to detailed information about UI elements and their associated functionalities.


\subsection{Guiding code generation}
\label{sec:results_guide}
Six out of 10 participants either agreed or somewhat agreed that \system allowed them to guide the LLM to generate code. Participants agreed that the ephemeral UIs helped them reduce the effort of excessive prompt engineering and scaffolded them to customize code generation. They also pointed out that while ephemeral UIs are helpful in many ways, there could be scenarios where text-based prompts could lead to desired code generation outcomes more effectively, suggesting a hybrid approach for users to flexibly choose UI or text-based approaches to interact with LLMs.

\subsubsection{Reducing effort of prompt engineering}
Participants thought \system reduced the effort of prompting compared to existing code generation tools with which they commonly needed to write detailed text prompts. For example, P7 shared their frustration of excessive prompt engineering: ``I need to give ChatGPT very detailed prompts to make it work. It's very frustrating'' (P7).
In contrast, ephemeral UIs introduced a scaffold of ``templating prompts,'' (P3) which added a UI layer where users could specify details about the code to be generated that they might initially omit. For instance, P6
found it straightforward to select from a color picker within the UI when requesting the LLMs to generate visualization code: ``it's allowing you to define those variables that you wouldn't otherwise be able to define easily.'' Users appreciated how ephemeral UIs made it easier for them to articulate details for code generation, allowing them to specify aspects of the code they would not be able to in textual prompts.

\subsubsection{Scaffolding customization in code}
\system makes areas where users can customize their code more apparent. P1 appreciated the ephemeral UIs for ``pulling out the things that you can configure, and helping you understand what are the parts of the code you actually have to customize.'' Users were able to realize the decisions needed to be made before code generation: ``when you generate the code, you can get that information upfront. This is the stuff that you need to configure, this is what it does, and this is why it matters. Rather than after the code is generated and then you got to do research'' (P6.) Ephemeral UIs facilitate users to thoughtfully guide code generation rather than passively receiving decisions made by LLMs. For example, P2 highlighted that while with ChatGPT she usually followed what the LLM suggested, with \system she could easily tailor the approaches: ``(\system) gives you more control of what kind of model structure you prefer.'' In summary, ephemeral UIs encourage users to proactively identify areas in programming where they need to make decisions---and assist with their decision-making.





\subsubsection{Opportunity---allowing users to choose UI or text to interact with LLMs}
While participants generally agreed that our system scaffolded their code generation process, some described situations where they would like to prompt the code generation in text. UIs are recognized as beneficial for learning and exploratory tasks, while text-based prompt approaches are suited for determined goals. As P9 pointed out, ``if I'm writing code that I have a pretty good idea of what I want to write and I just need to look up the syntax, I will prompt LLMs with text. I think when I'm learning, it (the ephemeral UI) allows for a little bit more exploration embedded into the learning.'' Such comments suggest the opportunities for a hybrid approach where both kinds of tools can be used in different phases of the code generation process. 

\begin{figure*}[ht]
\centering
  \includegraphics[width=\linewidth]{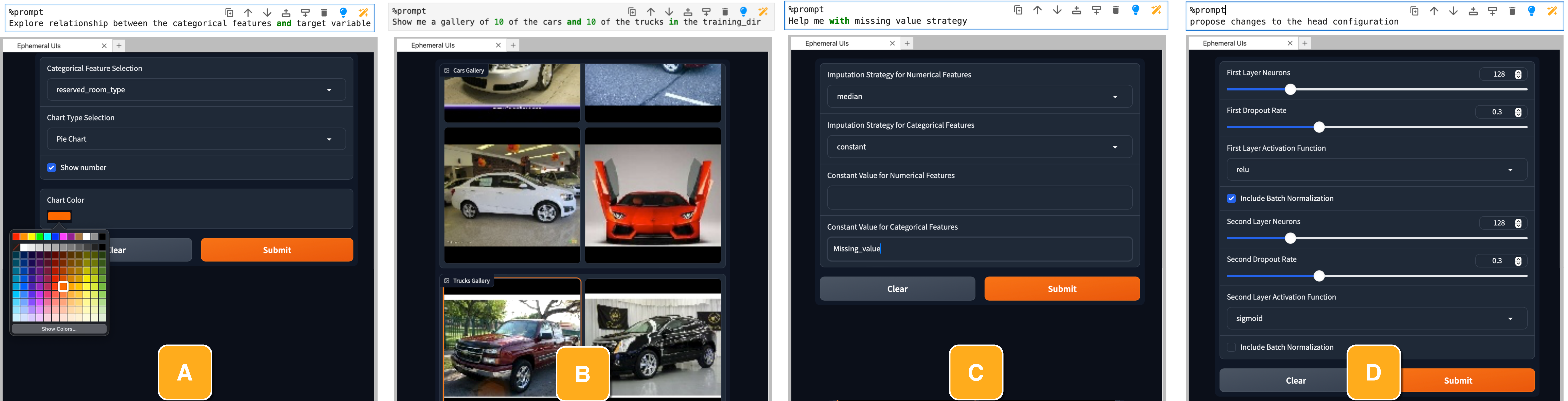}
  \caption{Examples of user request and ephemeral UI in our study. \protect\ovcircle{A}: UI generated by P8 for plotting. \protect\ovcircle{B}: UI generated by P4 to view training data. \protect\ovcircle{C}: UI generated by P5 to impute missing values. \protect\ovcircle{D}: UI generated by P3 to construct a model. }
  \label{fig:user_study_examples}
\end{figure*}

\subsection{Exploring code}
\label{sec:results_exploration}
Eight out of 10 participants either agreed or somewhat agreed that \system supported them when exploring machine learning methods in the tutorials. Participants reported on how \system supported exploration through facilitating iteration on existing examples and offering inspirations for new directions. At the same time, some participants pointed out that merely providing options in the UIs alone might not be sufficient for users to effectively iterate on their design, as some machine learning parameters were often determined by outcomes later in the workflow. They shared ideas on how to further incorporate guidance and rapid feedback into \system.


\subsubsection{Facilitating iteration on existing examples}
\system enabled users to efficiently and effectively iterate on different variations of existing code in the tutorials. For example, with \system, participants were able to experiment how choices on different machine learning parameters affected classification results: ``it (the dropdown menu for label modes) allows me to do experiments, changing from \texttt{categorical} to \texttt{int} or \texttt{binary}. And you could see how that affects the model later on'' (P9). Compared to manually editing code or prompting LLMs in text, ephemeral UIs were regarded as more effective because simple interactions with UI elements generated new versions of the code:  
``Sometimes in ChatGPT you have to write the prompt again, it will type everything and you will copy and paste. (With \system,) let's say I had chosen one category, but I actually wanted to see another one. I can just click it and I can go back and forth. This is way better from a iteration point of view'' (P8).

Ephemeral UIs also offer a playground for exploration that is separate from the main workflow.
Exploratory activities, such as viewing data samples, are encapsulated within the UIs, thereby not impacting the main workflow. 
As P4 described, ``if I wanted to generate a random selection of 10 images, (without \system) I would need to create a random string or dictionary, and then open them myself. This (\system) is just way simpler and saves me from code that I don't have to write or ask LLMs to write'' (\Cref{fig:user_study_examples} \ovcircle{B}). 


\subsubsection{Getting inspirations for new directions}
Participants shared that \system illuminated various options for machine learning variables previously unknown to them. 
Surfacing the available options can inspire users to expand the original example in the tutorial. For example, P10 appreciated the various options displayed on the ephemeral UIs: ``sometimes in ChatGPT, my impression is that if I specify one approach, it follows that path with little room for deviation or improvement. In contrast, here, without me having to ask explicitly, I'm presented with three to four options.'' That is, the UIs broaden the horizon for participants and encourage creativity.

Being able to view different options also aids users in reflecting on the nuances between various choices. For instance, the imputation options offered by an ephemeral UI prompted P10 to think about how the different strategies might impact data differently: ``it gave me the option of using the \texttt{mean}, \texttt{median}, or \texttt{mode}. I might have defaulted to using the mean, but thanks to the UI, I started questioning my choice. Why did I choose mean? Why not median? This level of engagement is definitely beneficial for learning.'' These examples highlight how \system can help users engage in critical thinking and decision-making through facilitating exploration.

\subsubsection{Opportunity---enhancing exploration with guidance and rapid feedback}
Some participants noted the importance of receiving feedback to navigate their exploration with the system effectively. In certain instances, it might not be feasible for users to select specific values in UI elements immediately---as these values often require experimental determination. For instance, when adjusting dropout rate in a model using sliders, P1 commented, ``because the dropout is typically a parameter that you would run different experiments to decide, I don't know if I would necessarily need this to set those values at this early stage.'' 
Similarly, P2 wished to preview the final classification results in order to choose from the imputation strategy dropdown menu: ``I would certainly be curious about how each of my choices affected the result.''
Therefore, the ephemeral UIs on their own might not be sufficiently helpful; users require guidance on how to interact with the UIs from effective and dynamic feedback deriving from future steps in the machine learning workflow.

\subsection{Efficiency in working with tutorials}
In our study, 9 out of 10 participants either agreed or somewhat agreed that our system enhances their efficiency while working through machine learning tutorials. This finding was somewhat counterintuitive, considering our system introduces an additional step---interacting with ephemeral UIs---into the code generation workflow. Participants provided several reasons \system helped them work with tutorials efficiently.

\subsubsection{Producing code templates}
Even experienced Python programmers found it helpful that the UIs could provide them with code templates. For instance, P2 noted how the ephemeral UIs offered a customizable starting point for function generation: ``It's a much more intuitive way to experiment with all the different inputs and properties without having to manually type everything out in Python'' (P2). 
Similarly, P9, who disliked writing or editing code for visualization, found that the ephemeral UI (\Cref{fig:user_study_examples} \ovcircle{A}) made variables in visualizations easily accessible: ``I can simply choose [from color pickers] to make this red and make this blue. Plotting code is perhaps one of my least favorite tasks. The UI is really useful here because it allows you to specify things visually, and it generates a significant block of code for you'' (P9). 

\subsubsection{Providing in-context support}
Participants appreciated that the ephemeral UIs were ``embedded''  (P9) in the programming environment of JupyterLab, offering them in-context support.
In particular, ephemeral UIs helped users navigate unfamiliar syntax and save time from searching documentation: ``I think without the help of this UI, I wouldn't have been able to get through the tutorial nearly as quickly. To fill out the optimizer part, I would have to go to the docs and look up the compiler syntax and try to find where I'd find those versus in the UI where it just filled it out for me'' (P1).

Additionally, in-context support from the UIs reduced distraction from the main programming workflow: ``I don't have to change tabs. I don't have to set a break to my flow or have multiple windows open to do something'' (P9). 

\subsubsection{Opportunity---opting in ephemeral UIs based on user preference}
P2 was the only participant who somewhat disagreed that our system enhanced their efficiency in machine learning tutorials. Comfortable with Python programming, P2 found that the ephemeral UIs could be ``getting in the way'' when she already had a clear idea of the program she wanted to write. She highlighted a preference for bypassing the ephemeral UIs and directly editing the code herself in such instances. This feedback connects to the findings in \Cref{sec:results_guide} indicating that users appreciate the flexibility to employ UI and text for different programming tasks.


\section{Discussion}
Our user studies revealed that users' experiences align closely with our Design Goals. We now delve into the broader implications of LLM-generated ephemeral UIs in code generation beyond the specific context of machine learning tutorials. 

\subsection{UI-centric interaction with code generation LLMs}
We introduced the workflow of ephemeral UIs, a novel interactive paradigm for code generation. Different from existing code generation tools such as ChatGPT and GitHub Copilot, our workflow dynamically generates UIs catered to users' intentions and the existing code context, and contributes to the ongoing discussions around how
LLMs bring up new opportunities for UI-based interactions. For example, Jovanovic~\cite{Jovanovic_2024} brought up the concept of ``generative UIs''---LLM-powered interfaces that dynamically cater to specific user needs. Moran and Gibbons~\cite{Moran_2024} devised the design framework of ``Outcome-Oriented Design'' that focuses on ``user goals and final outcomes, while strategically automating aspects of interaction and interface design.''  Our work demonstrates that ephemeral UIs offer affordances adapted to users' evolving intentions in programming. We believe that the UI-centric workflow has the potential to influence a variety of programming environments. 


\subsection{Ephemeral UIs for exploratory programming in notebooks}
We found that ephemeral UIs facilitate exploratory activities in computational notebooks. Programmers commonly engage in exploratory programming activities that are not directly relevant to their main workflow~\cite{kery2017exploring, kery2018story}. However, the linear cell structure designed for a sequential flow of code in computational notebooks often clashes with the convoluted and iterative processes of exploration~\cite{rule2018exploration, chattopadhyay2020what}. 
Existing research on notebooks has explored designs that facilitate better management of code variations and explorations within notebooks~\cite{kery2018interactions, rule2018aiding, kery2019towards}.
Our approach adds to this body of work by offering dynamically generated UIs that scaffold exploration. Ephemeral UIs address the challenges of exploration by containing certain exploratory steps within the UI, mitigating the messiness in the notebook by streamlining and reducing the exploratory code. We also observed users explore various options for function properties through dropdown menus, think about their implications on code in advance, and thereby make informed decisions about the code to be generated. This suggests that ephemeral UIs can incorporate understanding with exploration, easing the tension between the two~\cite{rule2018exploration} by prompting thoughtful reflection with code in action~\cite{schon2017reflective}. 

User feedback highlighted in \Cref{sec:results_exploration} indicated that merely providing UIs may not be sufficiently helpful. Besides the UIs, users need effective and dynamic feedback to inform their exploration. Future research could explore mechanisms to present users with previews on the implications of their interaction with the UIs. For example, interfaces could incorporate predictive models that simulate the outcomes of different user choices within the UI to provide feedback and integrate contextual help or tooltips that explain the potential impact of each option based on user data or best practices. 

\subsection{Introducing dynamic and in-situ scaffolds to tutorials} 
In line with Moran and Gibbons~\cite{Moran_2024}, our work explores the potential of serving ephemeral UIs in programming tutorials. Traditional UI supports---although carefully and aesthetically designed to afford complex user activities---are pre-defined and typically locked into the initial assumptions and design choices made by software authors. There is a cost both for authors to write the UIs and for users to learn to use those UIs.

In contrast, ephemeral UIs are automatically generated and dynamically adapted to the code context as users engage with programming tasks. This flexibility allows for a more personalized and in-situ learning experience, giving users relevant support. The dynamic nature of the UIs also means that users are presented with a cleaner interface instead of complex navigational paths, lowering the barrier to entry for users by eliminating the need to master complex software.

Although the ephemeral UIs generated in our particular prototype are relatively simple and based on the Gradio API, our work demonstrates the potential for LLMs to generate UIs that are useful for scaffolding users to explore data, to work with unfamiliar frameworks, and to iterate on their coding ideas.
We envision future development of a hybrid approach that combines pre-defined and dynamically generated UIs in tutoring systems. Some of the features might remain predefined to ensure stability and consistency, while others can be dynamically generated to provide personalized support tailored to specific coding episodes. 

\subsection{Limitations}
The usage of Gradio in \system limits the generated ephemeral UIs to those available in its library. 
Because our system is a JupyterLab extension, we were restricted to interactions that are possible within the JupyterLab interface. 
While we have gained qualitative insights on how users understand, guide, and explore code generation with \system and how users compare it with commercial tools such as ChatGPT, we did not test the findings quantitatively. 
We also recruited participants from a single organization. Although our organization does not own any technologies utilized in \system, participants can still have biases due to our shared affiliation. As a result, it is possible that participants were more reserved in providing critical feedback.
Finally, our study focused on users working with machine learning tutorials. The potential applications and use cases for LLM-powered ephemeral UIs extend beyond the domains of just programming tutorials, or even code generation more generally. We invite future researchers to further explore this space.

\section{Conclusion}

In this paper, we introduce a UI-centric workflow to LLM-based code generation tools that serves as an interface between text-based user prompts and the code to be generated. Through our system, \system, we introduce LLM-generated ephemeral UIs as a helpful assistant in machine learning tutorials. Our user study found that \system not only facilitates a deeper understanding of coding concepts but also encourages code exploration among users. The insights and implications from our research contribute guidance for the design and implementation of UI-centric experiences in LLM-powered code generation, with potential applicability in programming environments beyond computational notebooks. 
\bibliographystyle{IEEEtran}
\bibliography{reference}

\clearpage

\end{document}